\documentclass[12pt]{article}
\pdfoutput=1
\input{epsf}
\usepackage{epsfig}
\usepackage{amssymb}
\usepackage{amsfonts}
\usepackage{amsbsy}
\usepackage{eufrak}
\usepackage{amsmath}

\usepackage{color}
\usepackage{soul}

\newcommand{\be}{\begin{equation}}
\newcommand{\ee}{\end{equation}}
\newcommand{\ba}{\begin{array}}
\newcommand{\ea}{\end{array}}
\newcommand{\bea}{\begin{eqnarray}}
\newcommand{\eea}{\end{eqnarray}}

\newcommand{\nl}{\nonumber \\}

\def\G {{\cal G}}

\def\half{\frac{1}{2}}
\def\one{{\hbox{ 1\kern-.8mm l}}}

\def\ba{\bar{a}}

\def\({\left(}
\def\){\right)}
\def\[{\left[}
\def\]{\right]}
\def\d{\partial}
\def\hg{\hat{g}}

		\renewcommand{\baselinestretch}{1.4}
\begin{document}

		\begin{titlepage}
		
		\begin{flushright}
		YITP-SB-{1403}\\
		\end{flushright}
		\vskip.2in
		\begin{center}
		{\huge Time-Dependent Warping and Non-Singular}\\
		\vspace{.2cm}
		{\huge Bouncing Cosmologies}\\
		\end{center}
		\begin{center}
		{\large Koushik Balasubramanian, Sujan P. Dabholkar}\\
		\vspace{0.15cm}
		C. N. Yang Institute for Theoretical Physics, Stony Brook University, Stony Brook, NY 11794, USA\\
		{\tt koushikb@insti.physics.sunysb.edu}
		\end{center}
		\vspace{0cm}
		\begin{center}
		{\large ABSTRACT}
		\end{center}
		{
		In this note, we construct a family of non-singular time-dependent solutions of a six-dimensional gravitational theory that are warped products of a four dimensional bouncing cosmological solution and a two dimensional internal manifold. The warp factor  is time-dependent and breaks translation invariance along one of the internal directions. When the warp factor is periodic in time, the non-compact part of the geometry bounces periodically. The six dimensional geometry is supported by matter that does not violate the null energy condition. We show that this 6D geometry does not admit a closed trapped surface and hence the Hawking-Penrose singularity theorems do not apply to these solutions. We also present examples of singular solutions where the topology of the internal manifold changes dynamically.}

		\noindent
		\vspace{3mm}

		\end{titlepage}
		\newpage
		\renewcommand{\baselinestretch}{1.1}  
		\setcounter{page}{1}


\section{Introduction}

{\hspace{0.6cm}}
If the universe began with the big bang singularity, then it seems essential to find a ``theory of initial conditions". One approach to understanding the theory of initial conditions is to assume the existence of {the} universe before the big bang. It was shown that it is necessary to have a contracting phase for the universe to be past-eternal \cite{GuthBordeVilenkin}. Hawking and Penrose showed that a globally hyperbolic contracting space admitting a closed trapped surface will collapse into a singularity, unless an energy condition is violated \cite{Hawking}.
This {result} imposes severe restrictions on a smooth transition from a contracting phase to  an expanding phase. 
 
There are many phenomenological models incorporating a pre-big bang scenario, in which the singularity is avoided by having matter that violates {the} null energy condition (NEC) \cite{Bouncing,ghost} or by violating the NEC using modified gravity \cite{Modified}. For closed universes, it is sufficient to relax the strong energy condition (SEC) to avoid the singularity and such an example was constructed in \cite{SHU}.{\footnote{The Hawking-Penrose singularity theorem \cite{Hawking} assumes the strong energy condition to show the existence of singularities in closed universes.}} 

Though it is not possible to derive the energy conditions from first principles, it is known that most models of classical matter satisfy the NEC. Violation of the NEC implies that the Hamiltonian is unbounded from below. In other words, cosmological models violating the NEC admit solutions that have infinitely negative energies \cite{Vikman1}. It has also been argued that violation of NEC in certain models of ghost condensation are pathological due to the existence of superluminal instabilities \cite{Dubovsky}. However, there are models violating the NEC that do not admit modes that propagate with superluminal speeds \cite{Vikman2} or superluminal instabilities\cite{Vikman3}. It is not clear at this point if violation of such energy conditions is unphysical. The strong energy condition is violated by a positive cosmological constant and also during inflation. Relaxing strong energy condition seems benign. In this regard, it would be interesting to find a microscopic realization of the fluid stress tensor in \cite{SHU} using classical fields and a cosmological constant. Quantum effects can lead to violation of {the} null energy condition, but an averaged null energy condition must be satisfied. Orientifold planes in string theory can also allow for localized violations of {the} null-energy condition. 

{Instead of violating null and strong energy conditions, some researchers have sought to understand the singularity outside the realm of classical Einstein gravity.} For instance, there have been a large number of proposals in the literature to understand the initial singularity using string dualities \cite{BrandVafa}-\cite{KOSST}. We will now review some of these proposals briefly.

In \cite{vengasperini}, the cosmological solution is obtained by connecting two singular solutions at the singularity. A scalar field with a singular profile provides the stress tensor required to source the metric. 
Even though the infinite past is described by a smooth perturbative vacuum of string theory, the perturbative description breaks down near the bounce singularity and a non-perturbative string description is required to bridge the post big bang universe and the pre-big bang universe.

A geometric picture of certain big bounce singularities in higher dimensions was presented in \cite{KOSST, Seiberg}, where the lower dimensional scalar field uplifts to the higher dimensional radion field.{\footnote{ Also see \cite{GibbHorTown, Behrndt,larsenWilczek, Feinstein} for related work.}} The size of the circle shrinks to zero size when the universe passes through the singularity and expands again when the universe bounces from the singularity. They also considered the case where the compact direction is a line interval instead of a circle. In this case, when the universe approaches a big crunch, the branes at the endpoints of the interval collide with each other, and they pass through each other when the universe expands again \cite{KOSST, Seiberg}.{\footnote { This is slightly different from the original ekpyrotic model \cite{KOST}.}} The higher dimensional geometry discussed in \cite{KOSST, Seiberg} is simply a time-dependent orbifold of flat space-time. 

There are many Lorentzian or null orbifold models of bouncing singularities where the geometry is just obtained by taking quotients of flat spacetime by boost or combination of boosts and shifts \cite{Orbifold,john,LMS2}.{\footnote{ Since these geometries are locally flat, they are exact solutions of classical string theory.}} In the case of singular orbifolds, there is a circle that shrinks to zero size and then expands, leading to a bounce singularity. Such solutions are unstable to introduction of a single particle as the backreaction of the particle and the infinite number of orbifold images produces regions of large curvatures \cite{john, HoroPol}. In \cite{john, LMS2}, examples of non-singular time-dependent orbifolds were presented. In these examples, size of the compact directions remain non-zero at all times but it becomes infinitely large in the infinite past and infinite future. That is, the extra dimensions are initially non-compact and then go through a compactification-decompactification transition. These null-orbifolds are geodesically incomplete unless the anisotropic directions are non-compact.

In this paper, we present a new class of non-singular bouncing cosmological solutions that has the following features:
\begin{itemize}
\item[1.] These are classical solutions of Einstein's equations sourced by a stress-energy tensor that satisfies {the} null energy condition.{\footnote{Senovilla \cite{SenovillaDust} found non-singular inhomogeneous geometries sourced by a fluid satisfying the NEC. However, a classical field configuration that produces the fluid stress-energy tensor is not known.}}
\item[2.] The stress-energy tensor sourcing the metric can be realized by classical fields.
\item[3.] All non-compact spatial directions are homogeneous and isotropic.
\item[4.] These solutions can be embedded in string theory.
\end{itemize}
Demanding homogeneity and isotropy in all spatial directions (including compact directions) rule out the possibility of finding such geometries. In fact, it can be shown that the metric $ds^2 = -dt^2 + a(t)^2 d\vec{x}^2$ cannot exhibit a bounce (classically) unless {the} null-energy condition is violated \cite{Toumbas,Seiberg}.
Hence, it is essential to include anisotropy or inhomogeneity in the compact extra dimensions to find non-singular bouncing cosmologies. We show that a time-dependent warped metric of the following form can exhibit bouncing behavior (non-periodic as well as periodic):
\be ds^2 =  \left[  \left( -e^{2 A(t, \theta_{k})}dt^2 +e^{2 B(t, \theta_{k})} d\vec{x}^2 \right) \right] + e^{2C(t,\theta_{k})}g_{i j} d\theta^{i} d\theta^{j} + 2 \zeta_{i}(t,\theta_{k}) dt d\theta^{i} \label{Cosmology} \ee
More precisely, we find six dimensional solutions of Einstein-Maxwell-scalar theory in which the metric takes the form in (\ref{Cosmology}). Note that the non-compact directions are homogeneous and isotropic. The compact directions have finite non-vanishing size at all times. Most higher dimensional resolution of curvature singularities that have appeared in literature rely on reducing along a shrinking circle \cite{GibbHorTown,Behrndt, larsenWilczek, Feinstein}. Note that the higher dimensional geometry could be geodesically incomplete even if all curvature invariants are finite. In particular, the higher dimensional solution is geodesically incomplete if it satisfies the assumptions of the Hawking-Pensrose singularity theorems \cite{GibbHorTown}. 

We show that our solutions are geodesically complete and hence non-singular. We show our solutions evade Hawking-Penrose singularity theorems as the they do not admit a closed trapped surface. These geometries are homogeneous and isotropic along the non-compact spatial directions $\vec{x}$. This non-trivial six-dimensional  solution can be uplifted to a locally flat solution in 7-dimensions using an $O(2,2)$ transformation. This transformation provides a simple method for generating time-dependent warping.
We show that the six-dimensional solution does not admit a time-translation symmetry. 

In this paper, we also present an example of a class of solutions where the topology of the internal manifold changes dynamically. Note that we need atleast six-dimensions (3+1 non-compact directions and 2 internal directions) to see a topology change in the internal manifold. We present a class of six-dimensional solutions, where the topology changes from a genus one surface to genus zero surface. Such solutions do not have any simple four-dimensional description as the topology change involves mixing among an arbitrarily large number of Kaluza-Klein modes.{\footnote{We found it convenient to work with six-dimensional examples. But it is straightforward to generalize these solutions to higher dimensions.}} 

Rest of the paper is organized as follows: In the section \S2, we briefly review scale factor duality and $O(d,d)$ transformations. We present examples of some interesting solutions that can be generated from trivial solutions using dimensional reduction and $O(d,d)$ transformations. In section \S3, we use $O(d,d)$ transformations to generate six dimensional solutions of the form (\ref{Cosmology}) and show that these are geodesically complete as they do not admit closed trapped surfaces. In section \S4, we conclude with a discussion on the results of this paper. We also present a short discussion on singular solutions with internal manifolds that dynamically change topology. 

\section{{Dimensional reduction, scale factor duality and $O(d,d)$ transformations}}
 In this section, we will briefly review some solution generating techniques and also present a brief survey of some interesting solutions (in the literature) that can be obtained using these solution-generating techniques. 

\subsection*{\it 2.1 Generating non-trivial solutions from trivial solutions using Kaluza-Klein reduction}
We will now present an example which has appeared multiple times in literature (see for instance \cite{Seiberg,KOSST, Traschen}) to illustrate the utility of Kaluza-Klein reduction as a solution generating technique. We start with a flat metric written as a product of two-dimensional Milne universe and $\mathbb{R}^{d-1}$:
\be ds^2_{{\cal M}_2\times \mathbb{R}^{d-1}} = -dt^2 + t^2 dy^2 +d\vec{x}^2 \label{Milne},~\varphi = 0,~H =0 \ee
This is a trivial saddle point of the following action: 
\be S = \int d^{D+1}x\int d^d y \sqrt{-g}e^{-2\varphi}\( R +  4 \partial_\mu \varphi \partial^\mu \varphi - {1\over 12} H_{\mu\nu \rho}H^{\mu \nu \rho}\),  \label{StrinAct}\ee
We will now show that dimensional reduction along $y$ direction of ${\cal M}_2\times \mathbb{R}^{d-1}$ produces a non-trivial solution of the $d-$dimensional equations of motion. Using the Kaluza-Klein reduction ansatz, we can write the higher dimensional solution as
$$ds^2 = e^{2\alpha\sigma}ds^2_{E,d-1} + e^{2\beta \sigma} dy^2, $$ 
where $\sigma = {\beta^{-1}} \log |t|$; $ds^2_{E,d-1}$ is the lower dimensional line element in Einstein frame, and $$\alpha^2 ={1\over 2(d-1)(d-2)}, \beta = -\sqrt{{d-2\over 2 (d-1)}}  $$
The action in (\ref{StrinAct}) can be consistently truncated to the following Einstein-scalar action in lower dimensions:{\footnote{The lower dimensional action is a consistent truncation of the higher dimensional action if all solutions of the lower dimensional equations of motion can be uplifted to solutions of higher dimensional action.}} 
$$ S_{d} = \int d^dx\sqrt{-g}\(R - {1\over 2} \partial_\mu \varphi \partial^\mu \varphi \) $$
The lower dimensional solution is
\be ds^2_{E,d-1} = t^{2/(d-2)} \(-dt^2 + d\vec{x}^2 \),~\sigma = -{\sqrt{2(d-1)\over (d-2)}} \log |t| \label{ldMilne}\ee
Recall that the higher dimensional metric is just a special coordinate patch on $d+1$ dimensional Minkowski space-time. However, the lower dimensional solution is non-trivial and does not admit a time-like killing vector. In fact, the lower dimensional geometry has a curvature singularity. Though the curvature invariants of higher dimensional geometry are all finite, the spacetime is geodescially incomplete \cite{GibbHorTown}.
The above $d-$dimensional solution and the uplift to $d+1$ dimensional ${\cal M}_2\times \mathbb{R}^{d-1}$ has been discussed in \cite{KOSST,Seiberg,Traschen} already. 

It is also possible to generate solutions with a non-trivial geometry as the starting point instead of flat space-time. For instance, the Hawking-Turok instanton can be obtained by reducing a bubble of nothing in five-dimensions \cite{Garriga}. Using this trick, it is possible to generate magnetic or charged dilatonic solutions (black holes or expanding cosmologies) starting from known uncharged solutions \cite{ Traschen,Melvin,Peet,SenBH, HorBH}. 

Now, we will discuss a different uplift of the lower dimensional solution in (\ref{ldMilne}). The solution in (\ref{ldMilne}) can also be uplifted to the following solution of the higher dimensional equations of motion 
\be ds^2_{1} = -dt^2 +t^{-2}dy^2 + d\vec{x}^2,~\varphi = - \log|t|,~ H=0.  \label{SFDsol}\ee 
We will now show that the above solution is related to a particular solution of Belinsky-Khalatnikov type \cite{BK73}. 
Recall that the action in (\ref{StrinAct}) is not the Einstein frame action. The saddle point of the Einstein frame action is obtained by a Weyl rescaling of the metric. After shifting to Einstein frame, the solution is given by
\be ds^2_E = t^{4\over (d-1)} \( -dt^2 + t^{-2}dy^2 + d\vec{x}^2 \) \ee
After the coordinate redefinition: $t^2 = 2\tau, \vec{x} = \sqrt{2} \vec{X}$, the above solution becomes a special case of Belinsky-Khalatnikov solution \cite{BK73} (with $d=3$). In the new coordinates the solution takes the following form
\be ds^2_E = \( - d\tau^2 +  \tau^{2 p_1}dX_1^2 +\tau^{2 p_2} dX_2^2 + \tau^{2 p_3}dy^2 \),~\varphi = -{q\over \sqrt{2}}\log (2\tau) \ee
where $p_1=p_2 = 1/2,p_3=0,  q={1/\sqrt{2}}$. Note that $p_1+p_2+p_3=1$ and $p_1^2+p_2^2+p_3^2=1-q^2$. Belinsky and Khalatnikov \cite{BK73} found more general time-dependent solutions of the above form where $p_i$ and $q$ satisfy the same relation. 

The solution in (\ref{SFDsol}) is related to the solution in (\ref{Milne}) by an $O(d,d)$ duality transformation. When $d$ translationally invariant directions are compactified, the lower dimensional effective action obtained by dimensional reduction enjoys an $O(d,d)$ duality symmetry \cite{maharanaschwarz}. These transformations are generalizations of the Buscher transformations \cite{Buscher}. 
An $O(d,d)$ transformation maps a classical solution of the equations of motion to a different classical solution \cite{vengasperini}. This property is helpful in generating new interesting solutions from known solutions (even from trivial solutions). Let us consider the action of an $O(d,d)$ duality transformation on the following solution
$$ ds^2 = g_{ab}dx^adx^b+ G_{ij} dy^idy^j, ~\varphi =\varphi_0$$
where $\d_{y^i}$ is a Killing vector. 
The action of a general $O(d,d)$ transformation is given by
\be M =
 \begin{bmatrix} 
    G^{-1} & G^{-1}B \\
      BG^{-1} & G- BG^{-1}B \\
   \end{bmatrix} \rightarrow \Omega^{T}M \Omega, \label{OddTrans}\ee
where $\Omega$ is a $2d\times 2d$ $O(d,d)$ matrix {\it i.e.,} $\Omega$ satisfies the following condition:
   \be\Omega^T \begin{bmatrix} 
      0& \mathbb{I}_{d\times d} \\
      \mathbb{I}_{d\times d} & 0 \\
   \end{bmatrix}\Omega = \begin{bmatrix} 
      0& \mathbb{I}_{d\times d} \\
      \mathbb{I}_{d\times d} & 0 \\
   \end{bmatrix} \dot{=}~ \eta
   \ee
 The matrix $M$ is a symmetric $O(d,d)$ covariant matrix. It is possible to write the action in a manifestly $O(d,d)$ invariant fashion using the double field theory formalism (see \cite{OlafBarton}). In the double field theory formalism, $O(d,d)$ transformations can be written as a generalized coordinate transformation of the generalized metric $M$. Note that when $\Omega = \eta$, $M \rightarrow M^{-1}$ which is a generalization of the scale factor inversion.
 
Scale factor duality (SFD) transformation is a special case of an $O(d,d)$ duality transformation (with $H=dB=0$).
When $dB =0$, the action of scale factor duality can be written as follows
$$ G_{ij} \rightarrow \tilde{G}_{ij}' = G^{-1}_{ij}, \quad \varphi\rightarrow\varphi '= \varphi_0 -{1\over 2} \log\(\mbox{det} G\), \quad H\rightarrow H=dB=0$$
Scale factor duality maps an expanding universe to a contracting universe. This forms the basis for the pre-big bang scenario of \cite {vengasperini}. Note that the solution in (\ref{SFDsol}) is related to the locally flat solution in (\ref{Milne}) through a SFD transformation.

In the next section, we will show that the solution generating techniques discussed in this section can be used to find non-singular bouncing cosmologies that do not admit any closed trapped surface.

\section{Non-singular Bouncing Cosmological Solutions}

\subsection*{\it 3.1. Solution of six dimensional Einstein-Maxwell-Scalar theory}
 
In this section, we will describe a method to obtain six-dimensional non-singular cosmological solutions with time dependent warping. The basic idea is to use a non-trivial parametrization of flat space that would produce non-trivial solutions after dimensional reduction or $O(d,d)$ transformations. We begin by writing down a line element for a flat metric in seven dimensions (with 3 non-compact spatial directions and 3 compact directions): 
$$
ds_{7}^{2}=-dt^{2}\left(1-r'(t)^{2}\right)+d\vec{x}^2+r(t)^{2}d\theta^{2}
+g_{\phi \phi}d\phi^{2}+ \left(\alpha^{2}g_{\phi\phi}+\beta^{2}\right)dz^{2}\nl
$$\be\qquad \qquad \qquad \qquad \qquad  +2\beta\cos\theta r'(t)dt dz  -2\beta\sin\theta r(t)d\theta dz+2\alpha g_{\phi \phi}d\phi dz \label{CrazyFlat}
\ee
where $g_{\phi\phi} = (R+r(t)\sin\theta)^{2}$; $\vec{x}$ denotes the 3 non-compact spatial directions, $t$ denotes a timelike coordinate, $\theta$, $\phi$ and $z$ are the 3 compact directions; $\alpha$, $\beta$ and $R$ are non-zero constants. Note that the metric degenerates when $\beta=0$. To ensure that $t$ is timelike, we choose $r(t)$ such that $-1<r'(t)<1$. The above metric can be transformed to the familiar flat space metric: $ds^2=-dt'^2 + d\vec{x}'^2 + d\vec{y}^2$, by using the following change of coordinates
$$ \vec{x}' = \vec{x}, ~t' = t, ~y_1 = \beta z + r(t) \cos \theta,~y_2=(R+r(t)\sin\theta)\cos(\phi +\alpha z),$$ \be~y_3=(R+r(t)\sin\theta)\sin(\phi +\alpha z) \label{eq:CoordTrans}\ee
with $-\infty > t > \infty$, $2\pi>\theta\ge 0$ and $2\pi>\phi\ge0$. {The Jacobian of these coordinate transformations vanish at $\theta = \pi/2$. However, this is just a coordinate singularity and not a physical singularity. We will prove that the geometry is not singular by proving geodesic completeness in a later subsection.} The metric in (\ref{CrazyFlat}) extremizes the seven dimensional low-energy string effective action in (\ref{StrinAct}) (with $\varphi = 0$ and $H=0$).
We will now reduce along $z$ direction to obtain a non-trivial solution in six-dimensions. The six dimensional action can be obtained by writing the 7D line element in the Kaluza-Klein reduction ansatz: 
$$ds^{2}_{7} = e^{-\sigma/2}d\hat{s}^{2}_{6}+e^{2\sigma}\left(dz+\hat{A}_{\mu}dx^{\mu}\right)^2.$$
 When $\varphi$ and $H$ are trivial, the seven dimensional action can be consistently truncated to the following Einstein-Maxwell-scalar action:
\be S^{(6)}_E =  \int d^{6}x \sqrt{-\hat{g}}\( \hat{R} - {5\over 4}\partial_\mu \sigma \partial^\mu \sigma- {1\over 4}e^{{5\over 2}\sigma} \hat{F}_{\mu\nu}\hat{F}^{\mu \nu}\),  \label{EMD}\ee
where, $\hat{g}$ is the Einstein frame metric, $\hat{F} =d \hat{A}$ is the field strength and $\sigma$ is the radion field. The six dimensional solution is given by (see appendix A)
$$
e^{2\sigma}= \alpha^{2}g_{\phi\phi}+\beta^{2},~$$
$$
\hat{g}_{tt}=-e^{{\sigma \over2}}\left(1-r'(t)^{2}\right) -\beta^{2} r'(t)^2e^{-{3\sigma\over2}}\cos^{2}\theta ,~
\hat{g}_{t\theta}=-e^{5\sigma/2} \hat{A}_t\hat{A}_\theta,~
\hat{g}_{t\phi}=-e^{5\sigma/2} \hat{A}_t\hat{A}_\phi,~
$$
\be
\hat{g}_{\theta\theta}=r(t)^{2}e^{{\sigma \over2}}\left(1-\beta^{2}e^{-2\sigma}\sin^{2}\theta \right),~
\hat{g}_{\theta\phi}=-e^{5\sigma/2} \hat{A}_\theta\hat{A}_\phi,~
\label{sol6d}
\ee
$$\hat{g}_{\phi\phi}=\beta^2 e^{-3\sigma\over2} g_{\phi \phi},~
\hat{g}_{ij} = e^{\sigma\over2}\delta_{ij}
$$
$$
\hat{A}_{t}={\beta r'(t)\cos\theta}e^{-2\sigma},~
\hat{A}_{\theta}=-e^{-2\sigma}\beta r(t)\sin\theta,~
\hat{A}_{\phi}=e^{-2\sigma}\alpha g_{\phi\phi}
$$
Other components of the gauge field and the metric are trivial. This six dimensional solution describes a $\mathbb{T}^2$ fibered over $\mathbb{R}^{3,1}$. Note that the metric on the $\mathbb{T}^2$ is not flat. The above solution can be uplifted to a different classical solution of a 7D theory described by (\ref{StrinAct}). This non-trivial solution is related to the trivial seven dimensional solution in (\ref{CrazyFlat}) by an $O(2,2)$ transformation (Buscher transformations). The details of this solution can be found in appendix B (see \ref{7Dsola} and \ref{7Dsolb}). Note that the 7D solution is regular if the six-dimensional solution is regular. 
The six-dimensional solution can be regular only if the size of the compact directions do not shrink to zero size. This is ensured by choosing $r(t)$ such that $R>r(t)>0$ for all $t$, and $\beta > 0$. With these conditions, the components of the metric and inverse metric are regular everywhere. All derivatives of the metric are also regular everywhere. All curvature invariants can be built from product of the derivative of metric components and inverse metric. Since the metric, inverse metric and their derivatives are all regular, all curvature invariants are finite. However, finiteness of curvature invariants does not imply the geometry is free of singularities. In order to show the six-dimensional solution in (\ref{sol6d}) is non-singular, we have to prove that it is geodesically complete \cite{Geroch}. 
We will prove this at the end of the next sub-section.

\subsection*{{\it 3.2. Absence of time-translation symmetry}}

In this subsection, we will show that our solution in (\ref{sol6d}) does not admit a time-translation symmetry. {By proving the absence of time-translation symmetry we also prove that it is not possible to get rid of the arbitrary function $r(t)$ in the solution by a gauge transformation, which can also be infered from the fact that the gauge field strength is non-zero.} In the process of showing this, we found a simple trick to prove our solution is geodesically complete. We will present this discussion at the end of this sub-section. 

We begin with a discussion on time translation symmetry. $\xi$ is a symmetry generator if the following equations are satisfied
\be \delta_\xi \sigma = \xi^\mu\partial_\mu \sigma = 0,\quad \delta_\xi \hat{A}_\mu = \xi^\nu \partial_\nu \hat{A}_\mu + \partial_\mu\xi^\lambda \hat{A}_\lambda = \partial_\mu \Lambda, \quad \delta_\xi \hg_{\mu \nu} = \nabla_\mu \xi_\nu + \nabla_\mu \xi_\nu =0 \label{KillEq}\ee 
where $\Lambda$ denotes the gauge shift. We can rewrite the second condition as follows:
$$ \xi^\lambda \(-\partial_\mu \hat{A}_\lambda + \partial_\lambda \hat{A}_\mu\)= \partial_\mu \Lambda -\partial_\mu \(\hat{A}_\lambda \xi^\lambda\) $$
\be \quad \delta_\xi \hat{A}_\mu = \xi^\nu \hat{F}_{\nu \mu} = \partial_\mu \tilde{\Lambda} \ee 
where $\tilde{\Lambda} = \partial_\mu \Lambda -\partial_\mu \(\hat{A}_\lambda \xi^\lambda\)$ is just a redefinition of the gauge shift.

We will now show that there is no time-like vector satisfying the above conditions. Note that $\xi^t$ must be non-trivial for $\xi$ to be time-like. The first two conditions and the trace of the third condition implies that $\xi$ should take the following form
$$ \xi = {U_0 \over \sqrt{\hg}}\(\partial_\theta \sigma \partial_t - \partial_t \sigma \partial_\theta + {\hat{F}_{t\theta} \over \hat{F}_{t\phi} } \partial_t \sigma \partial_\phi \) + {1\over \sqrt{\hg}} {\partial_t \(\tilde{\Lambda} (\sigma)\)\over \hat{F}_{t \phi}} \partial_\phi +V_0^i \d_i$$
where $\tilde{\Lambda}(\sigma)$ is a function of $\sigma$, $U_0$ and $V^i_0$ are constants. Note that we have used the isotropy and homogeneity of the non-compact spatial directions to write down the above expression. The variation of $\sigma$, $\hat{A}_\mu$ and the trace of the Killing equation seems to fix $\xi$ uniquely unto some unknown constants and an unknown function ($\tilde{\Lambda}$) of $\sigma$. The only freedom in $\xi$ is in the choice of $\tilde{\Lambda}$. The form of $\tilde{\Lambda}$ should be fixed by using the other Killing equations.  We can verify that there exists {\it no} $\tilde{\Lambda}(\sigma)$ for which $\delta_\xi \hg_{t\phi}$, $\delta_\xi \hg_{t\theta}$, $\delta_\xi \hg_{\theta \phi}$, $\delta_\xi \hg_{tt}$ and $\delta_\xi \hg_{\theta \theta}$ all vanish when $U_0 \neq 0$. We also know that $\xi$ is not time-like if $U_0 = 0$. This implies that the 6D solution does not admit a time-translation symmetry.
Note that when $r(t)$ is periodic, the geometry is invariant under discrete time translation invariance. 

We will now show that the 6D geometry is geodesically complete for any choice of $r(t)$ satisfying the conditions$:$ $0 < r(t) < R ~\forall ~t,$ and $\beta > 0$. To show this, we will first construct a vector $\zeta$ that satisfies $\nabla_\mu \zeta_\nu + \nabla_\nu \zeta_\mu =0$, but $\delta_\zeta \sigma \neq 0$. Note that such a vector is not a symmetry of the theory. For instance, linear dilaton solutions ten-dimensional supergravity theories admit such a vector \cite{CKR, ABKS}. In the linear dilation solutions, translation invariance {(along a particular direction)} is manifestly broken by the dilaton, while the string frame metric is invariant under spatial translations.{\footnote{Also see \cite{Qlattices} for an example of a solution of where translation invariance is broken by a complex scalar field, but not by the metric.}}

We will now return to our discussion on geodesic completeness. We can verify that the  $\zeta_\mu = e^{\sigma/2} \delta^0_\mu$ satisfies  $\nabla_\mu \zeta_\nu + \nabla_\nu \zeta_\mu =0$ but,
$$ \delta_\zeta \sigma = \zeta^\mu \d_\mu \sigma = 2 e^{-\sigma/2} {\alpha^2 r'(t) (\beta^2  + \alpha^2 r(t)^2) \sec \theta (R+ r(t)\sin \theta) \tan \theta \over \beta^2 r(t)^2} \neq 0.$$
We would like to emphasize that $\zeta$ does not generate time translation symmetry. However, the existence of this vector simplifies the proof of geodesic completeness. Let $u^\mu$ denote the tangent vector to a geodesic and $\lambda$ be an affine parameter. To prove geodesic completeness, we have to show that the affine parameter $\lambda$ can take all values in $(-\infty, \infty)$. Using the fact $\nabla_\mu \zeta_\nu + \nabla_\nu \zeta_\mu =0$ and the geodesic equation ($u^\mu \nabla_\mu u^\nu =0$), we can show that $u^\mu\zeta_\mu$ is a constant. This implies 
$$ {dt \over d\lambda} = \text{constant} \equiv E \implies \lambda =  {t\over E} + \text{constant}$$
We will now show that the derivative of $\theta$, $\phi$ and $\vec{x}$ with respect to the affine parameter are also finite. Translation invariance along $\phi$ and $\vec{x}$ implies 
 $$ g_{\phi \phi}{d\phi \over d\lambda} + g_{ \phi t}{dt\over d\lambda} +g_{\phi \theta}{d\theta \over d\lambda}  = L =\text{constant}, \quad e^{\sigma/2}{dx^i \over d\lambda} = p^i = \text{constant}$$
We alos know that $g_{\phi \phi}$ and $e^{\sigma/2}$ are non-zero and finite. Hence $d\vec{x}/d\lambda$ is also finite. We can write 
\be d\phi/d\lambda  = L - g_{\theta \phi}d\theta/d\lambda -g_{t \phi}dt/d\lambda \label{dphidlambda}\ee
We will now prove that the derivative of $\theta$ is finite. We will proceed by noting that $g_{\mu \nu} \dot{x}^\mu \dot{x}^\nu = k$, where $k=0$ for null geodesics and $k=-1$ for timelike geodesics. Hence,
\be
\hat{g}_{tt}\left(\frac{dt}{d\lambda}\right)^{2}+\hat{g}_{\theta\theta}\left(\frac{d\theta}{d\lambda}\right)^{2}+e^{-\hat{\phi}/2}\vec{p}^2+\hat{g}^{-1}_{\phi\phi}\(L - g_{\theta \phi}d\theta/d\lambda -g_{t \phi}dt/d\lambda\)^{2}+
2\hat{g}_{t\theta}\left(\frac{dt}{d\lambda}\right)\left(\frac{d\theta}{d\lambda}\right)
$$$$+2\hat{g}_{t\phi}\left(\frac{dt}{d\lambda}\right)\left(\frac{L - g_{\theta \phi}d\theta/d\lambda -g_{t \phi}dt/d\lambda}{\hat{g}_{\phi\phi}}\right)+2\hat{g}_{\phi\theta}\left(\frac{L - g_{\theta \phi}d\theta/d\lambda -g_{t \phi}dt/d\lambda}{\hat{g}_{\phi\phi}}\right)\left(\frac{d\theta}{d\lambda}\right)
=k 
\label{geodesic}\ee
where $\vec{p}$ and $L$ are conserved quantities associated with the $\vec{x}$-translation and $\phi$-translation Killing vectors: $\vec{\partial}$ and $\partial_\phi$. We have used equation (\ref{dphidlambda}) to eliminate $d\phi/d\lambda$ from the geodesic equation. We have already shown that $dt/d\lambda$ is a constant and hence finite. We also know that all components of the metric and $e^{\sigma}$ are bounded. Hence the geodesic equation (\ref{geodesic}) can be satisfied only if ${d\theta}/{d\lambda}$ is finite. Hence derivatives of $t$, $\theta$, $\phi$ and $\vec{x}$ with respect to the affine parameter are all finite.
This shows that $\lambda$ can take all values in $(-\infty, \infty)$ and hence the six-dimensional geometry is geodesically complete. In the next section, we show that our solution evades the Hawking-Penrose singularity theorem because it does not admit any closed trapped surface.

\subsection*{\it 3.3. Absence of Trapped Surface}

\begin{figure}[h]
\begin{center}  
\includegraphics[height=2.0in,width=4.40in,angle=0]{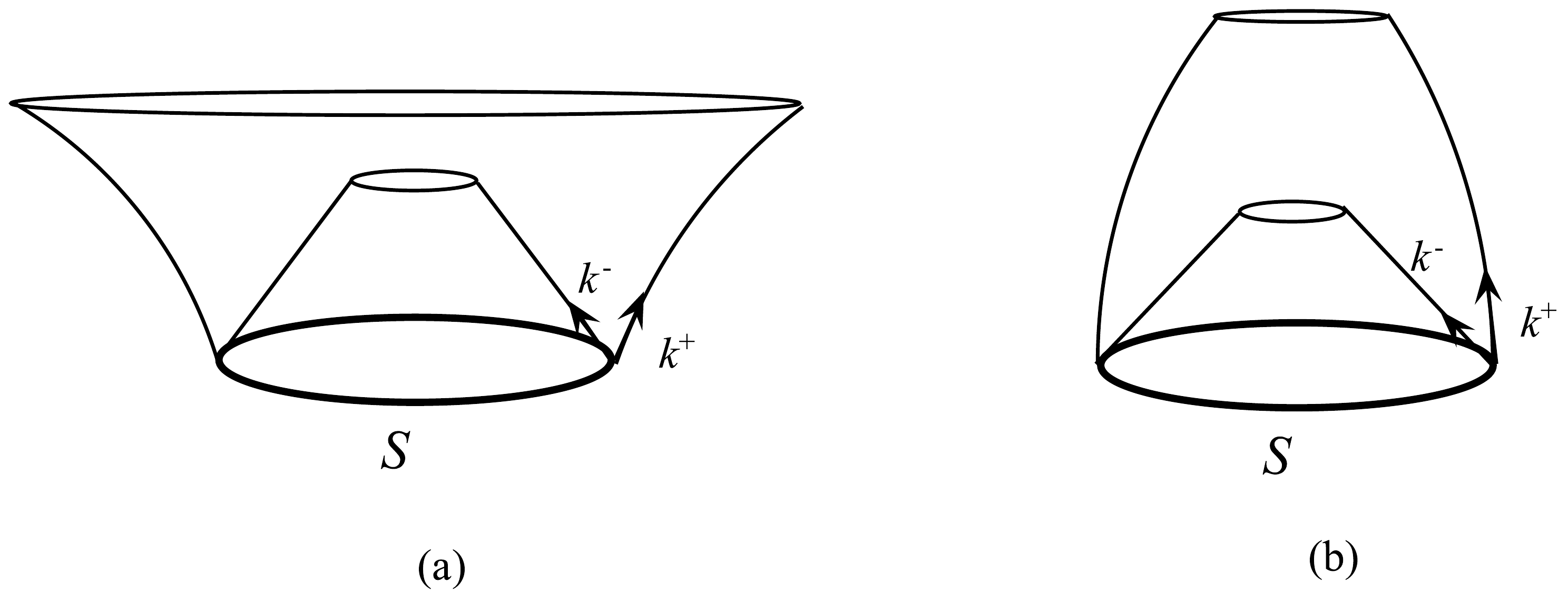}  
\caption{Shows  (a) an untrapped surface ($\kappa <0$) and (b) future trapped surface ($\kappa >0$). $k^+$ and $k^-$ are the null-vectors associated with the ingoing and outgoing null-congruences normal to the surface of $S$. \label{fig:Trap}}  
\end{center}  
\end{figure}

The existence of closed trapped surface (CTS) is an essential ingredient in the proof of Hawking-Penrose singularity theorems. A closed trapped surface is a compact codimension-two spacelike surface, where both ``ingoing" and ``outgoing" null-congruence normal to the surface are converging. In this section, we show that the geometry described by (\ref{sol6d}) does not admit such a trapped surface (see Fig. \ref{fig:Trap}). To prove the non-existence of CTS, we have to show that the product of the trace of the two null second fundamental forms is not positive. 

Before we proceed to the calculations, we will provide a simple argument for the non-existence of closed trapped surfaces in (\ref{sol6d}). 
The six dimensional solution is obtained by reducing (\ref{CrazyFlat}) along $z$ direction. The existence of a CTS in six-dimensions would imply the existence of a CTS in seven dimensions because a CTS in 6D (${\cal M}^{6D}_{CTS}$) will simply uplift to a CTS in seven dimensions (${\cal M}^{7D}_{CTS}~\equiv$ circle fibered over ${\cal M}^{6D}_{CTS}$) . But, the seven dimensional geometry does not admit a CTS since it is just a global coordinate patch covering entire flat space-time (which does not admit a CTS). Hence the six-dimensional solution does not admit a closed trapped surface. This argument relies on the fact that the size of the Kaluza-Klein circle is non-vanishing and finite.

 We will now show that the six-dimensional geometry does not admit a CTS by explicitly computing the product of the expansion factors. This also implies the non-existence of a CTS in seven-dimensions.
 First, we rewrite the six-dimensional metric in the following form for convenience.  
 \be
ds^2=-\hat{g}_{tt}dt^2 + e^{-\sigma/2} \left(d\rho^2 +   \rho^2 d\theta_{2}^{2}+ \rho^2 \sin\theta_{2}^2 d\phi_{2}^2 \right)+ 
2\hat{g}_{t\theta} dt d\theta + 2\hat{g}_{t\phi} dt d\phi $$
$$+ 2\hat{g}_{\theta \phi} d\theta d\phi  + \hat{g}_{\theta \theta} d\theta^2 + \hat{g}_{\phi \phi} d\phi^2
\ee
Since the non-compact spatial directions are homogeneous and isotropic, it is sufficient to show that a surface $S$, described by $t = t_0,~ \rho = \rho_0$ cannot be trapped, where $t_0$ and $\rho_0$ are some constants. The first fundamental form associated with the surface $t = t_0$, $\rho = \rho_0$ is 
$$ \gamma_{AB}dx^A dx^B =e^{-\sigma(t_0)/2} \left( \rho_0^2 d\theta_{2}^{2}+ \rho_0^2 \sin\theta_{2}^2 d\phi_{2}^2 \right) + 2\hat{g}_{\theta \phi}(t_0) d\theta d\phi  + \hat{g}_{\theta \theta}(t_0) d\theta^2 + \hat{g}_{\phi \phi}(t_0) d\phi^2$$
where $A, B \in \{\theta_2, \phi_2, \theta, \phi \}$. Note that this surface is a $\mathbb{T}^2$ fibered over a two-sphere. Now, we can define the future-directed ingoing and outgoing null 1-forms normal to this surface as follows
\be
k^{\pm} ~\dot{=}~ e^{\pm2\nu} e^{\sigma/4} (\frac{1}{\sqrt{2}},0,0,\pm \frac{1}{\sqrt{2}},0,0) 
\ee
where $\nu$ is an arbitrary function on the surface $S$. We can now compute the second fundamental form as follows:
\be \chi_{AB}^{\pm} = k^{\pm}_\mu \Gamma^{\mu}_{AB}{\Bigg|}_{S} = \displaystyle{{k_\mu^{\pm}  \hat{g}^{\mu \rho} \over 2} \left( \partial_{A} \hat{g}_{\rho B} +  \partial_{B} \hat{g}_{A\rho } -\partial_{\rho} \hat{g}_{AB} \right)}\Bigg|_S \ee
Now, let us define $\kappa = 2 \left({\gamma^{AB}} \chi^+_{AB}\right)\left( \gamma^{CD}\chi^-_{CD}\right)$. A simple procedure for computing $\kappa$ can be found in \cite{TrapSurf}. The product of the trace of  $\chi_{AB}^{+}$ and  $\chi_{AB}^{-}$ is given by
$$
\kappa =\Bigg[{r'(t_0)^2 e^{-17\sigma/2}\over \alpha ^6 \rho_0^4  {g_{\phi \phi}}^2 r(t_0)^2} \Bigg(\alpha ^3 \rho_0 ^2 \sin ^2(\theta ) r(t_0)^2 \left(2 {e^{4\sigma}}-\alpha  \beta  {e^{2\sigma}} \sqrt{{g_{\phi \phi} }}+2 \alpha  \beta ^3
   \sqrt{{g_{\phi \phi} }}\right)+\alpha ^3 {e^{4\sigma}} \rho_0 ^2 R^2 + $$ $${e^{2\sigma}} r(t_0) \Bigg(\beta  \left({e^{2\sigma}}-\beta ^2 \right)^2 \cos (\theta ) \cot
   ({\theta_2})+\alpha ^2 \rho_0 ^2 \sin (\theta ) \left(-2 \beta ^3+2 \beta  {e^{2\sigma}}+3 \alpha  R{e^{2\sigma}} \right)\Bigg)\Bigg)^2 - {4\over \rho_0^2 e^{\sigma/2}}\Bigg]_S$$Note that $\kappa$ is independent of $\nu$. We will now show that $\kappa$ cannot be positive everywhere if $S$ is compact ($S$ is compact only if $\rho_0$ is finite). First, note that when $r'(t_0)=0$, $\kappa$ is negative for all values of $\rho$. Hence, it is sufficient to consider the case where $r'(t_0)$ is non-zero. 
   
 Demanding positivity of $\kappa$ at $\theta = \pi$ we get,
$$\rho_0^2 > {{e^{-4\tilde{\sigma}}}\over  r'(t_0)^2}\Bigg[2 {e^{3\tilde{\sigma}}}{ r(t_0)^{3/2} \left({e^{2\tilde{\sigma}}} r(t_0)+\alpha  \beta  R^2 \cot ({\theta_2}) r'(t_0)^2\right)^{1/2}}+ $$
$$ 2 {e^{4\tilde{\sigma}}} r(t_0)^2+\alpha  \beta {e^{2\tilde{\sigma}}} R^2 \cot ({\theta_2}) r(t_0) r'(t_0)^2\Bigg]$$
where $e^{2\tilde{\sigma}} = \alpha^2 R^2 + \beta^2$. Note that when $\theta_2 \rightarrow 0$, $\rho_0 \rightarrow \infty$ ($\alpha$ and $\beta$ are non-zero). Similarly, $\rho_0$ diverges when $\theta_2 \rightarrow \pi$. Hence, $\kappa$ cannot be positive when $\theta =\pi$ and $\theta_2 = 0~\text{or}~\pi$ unless $\rho_0$ is infinite. This shows that a trapped surface cannot be compact and hence the 6D solution in (\ref{sol6d}) does not admit a closed trapped surface.
\section{Discussion}

In this note, we studied a family of six-dimensional (and 7D) nonsingular cosmological solutions that can be obtained from 7D flat spacetime using simple solution generating techniques. We have shown that our solutions are free of closed trapped surfaces and hence they evade the Hawking-Penrose singularity theorems. 
Since, these solutions can be generated from flat space, it is straightforward to embed these solutions in string theory. In particular, the 7D solutions in appendix B can be obtained from solutions of type II supergravity by reducing along a $\mathbb{T}^3$ (with all RR field strengths set to zero). 

In order to understand the physics as seen by a four dimensional observer, it seems essential to study the reduction to four-dimensions. 
However, it appears that the 6D and 7D solutions discussed in this paper do not have any simple description in four dimensions. When the warp factor is time dependent all Kaluza-Klein modes are excited and it is not clear how the higher Kaluza-Klein modes decouple from the lower dimensional effective action. There has been some work in the literature to understand the quadratic terms appearing in the lower dimensional effective action \cite{TWF} in a general warped compactification. But at this point it is not clear how one can study the non-linear terms arising from such a reduction. In a general warped compactification, the nonlinear terms lead to mixing between arbitrary number of Kaluza-Klein modes and a procedure for consistently truncating to the lowest Kaluza-Klein modes is not yet known. 

\begin{figure}[t]
   {\centering
   \includegraphics[trim = 10mm 10mm 10mm 10mm, clip=true, scale=0.4]{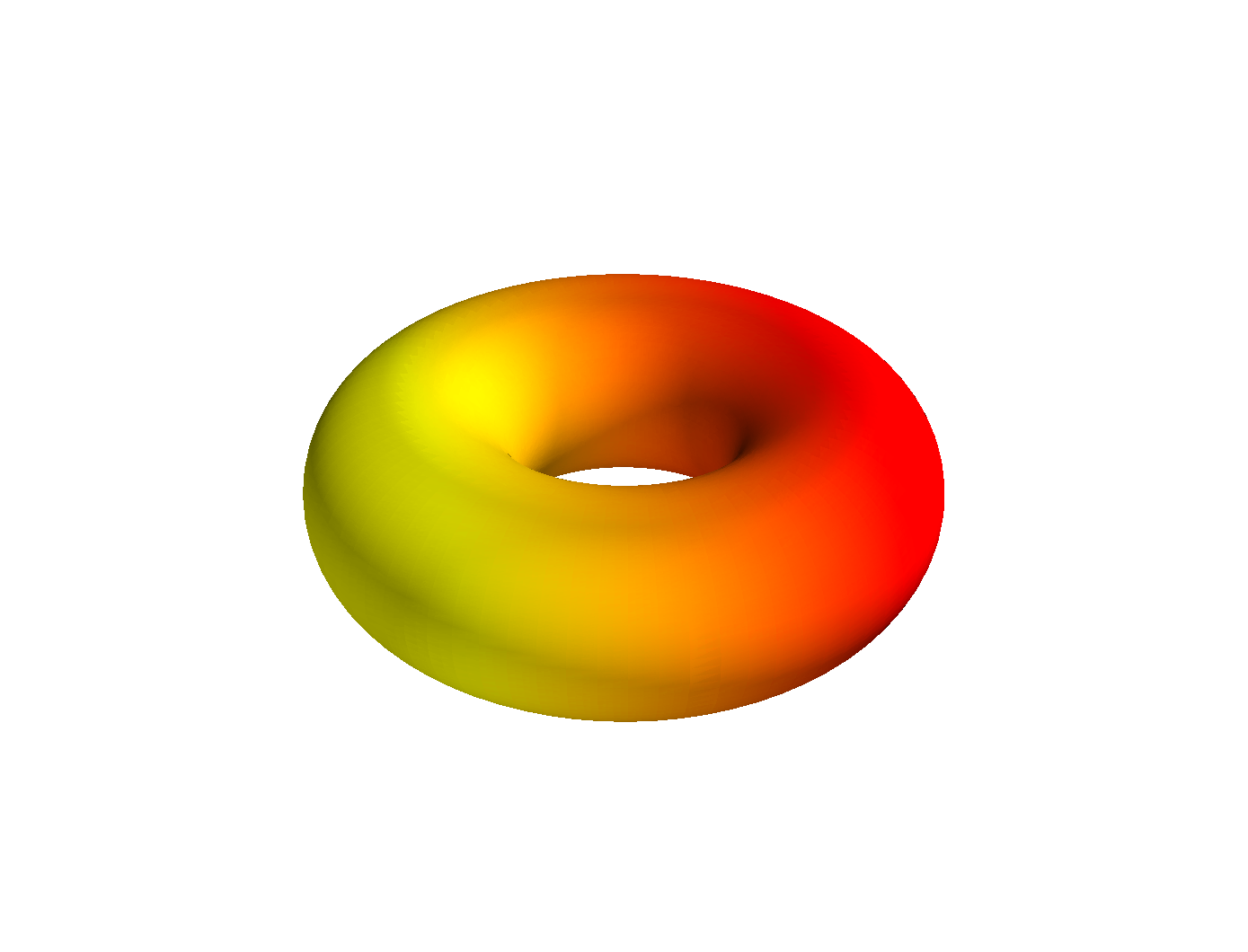}  
   \put(5,50){$\rightarrow$}
   \hspace{1cm}
      \includegraphics[trim = 20mm 10mm 10mm 10mm, clip=true, scale=0.4]{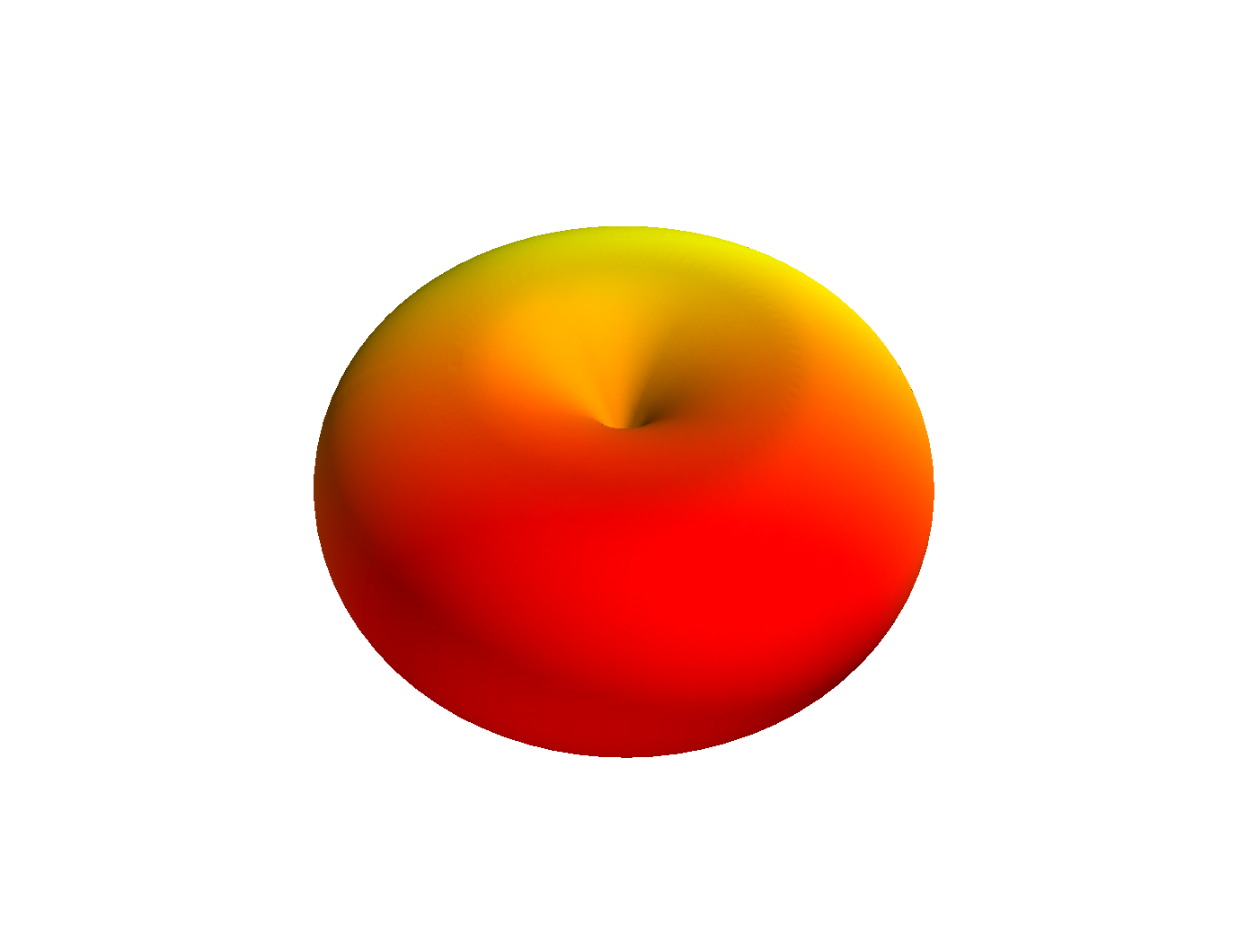} 
            \put(5,50){$\rightarrow$}
   \hspace{1cm}
         \includegraphics[trim = 20mm 10mm 10mm 10mm, clip=true, scale=0.4]{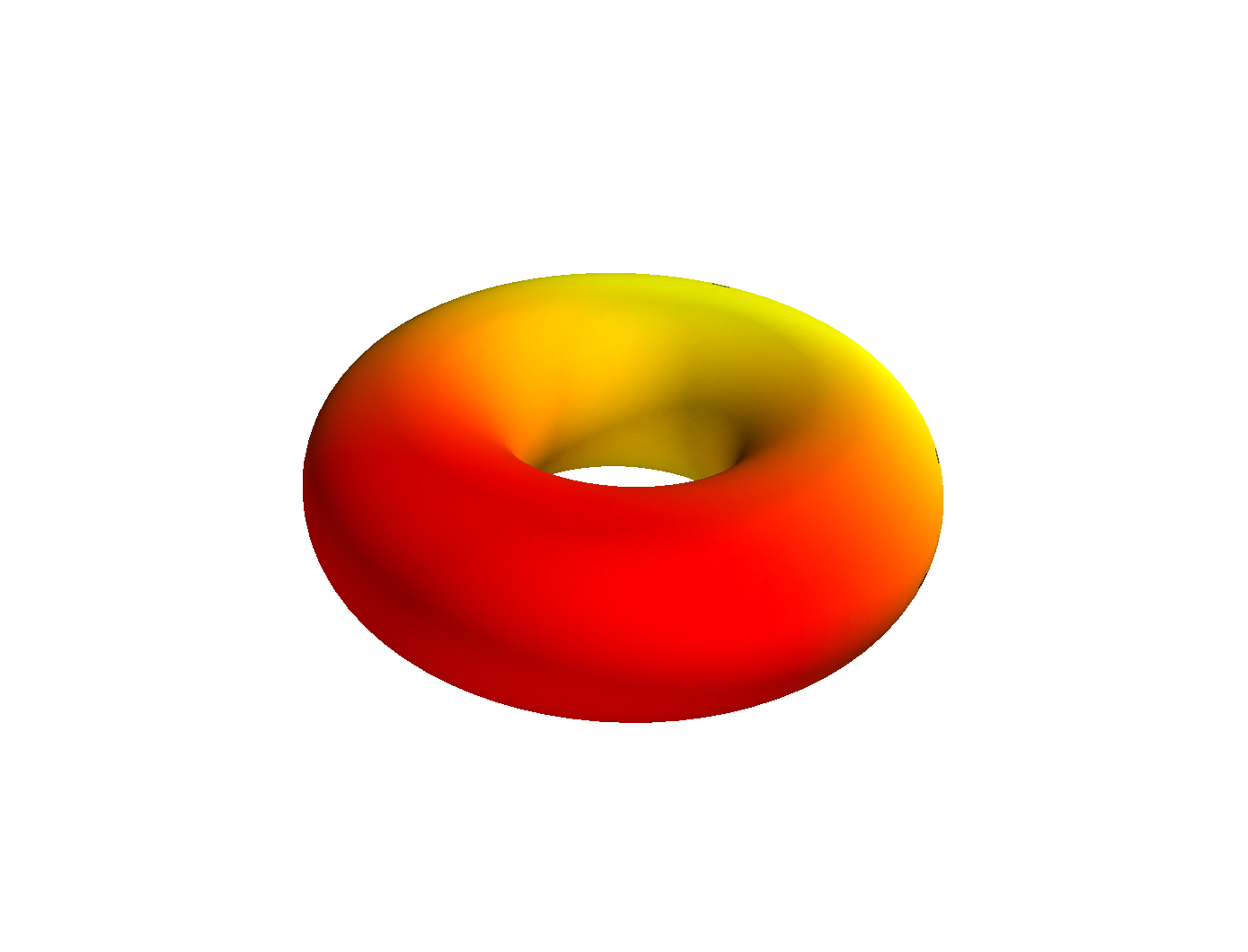} 
   \caption{Topology change from a surface with topological genus one to a surface with genus (topological) zero.}
   \label{fig:topchange}}
\end{figure}

In this note, we have only focussed on geometries that are warped products of a $\mathbb{T}^2$ and 3+1 dimensional bouncing cosmology. However, the method used to obtain theses solutions can be used to generate solutions where the topology of the internal manifold is different from $\mathbb{T}^2$. In fact, there are solutions where the topology of the internal manifold changes dynamically. We will provide a simple example of such a solution here. Let us consider the solution in (\ref{sol6d}) when $ \text{min} (r(t))< R \le \text{max} (r(t))$. When $r(t) < R$, the internal manifold is a ring torus and the six-dimensional metric in (\ref{sol6d}) describes a $\mathbb{T}^2$ fibered over $\mathbb{R}^{3,1}$, while the internal geometry has topological genus zero when $r(t) \ge R$ (see Fig. \ref{fig:topchange}). This topology change can also happen periodically if $r(t)$ is periodic. Such topology changing transitions are singular ($g_{\phi \phi}$ vanishes when $r(t) = R$) even though the scalar field and the gauge field strength do not diverge. The Euler characteristic of the internal manifold is zero even when $r(t) \ge R$ because of the singularities.{\footnote{ The Euler characteristic of a Riemann surface described by an algebraic curve with $N_s$ singular points of multiplicities $m_1,\cdots, m_{N_s}$ and topological genus $g$ is $$\displaystyle{\chi_e  = 2 - 2g - \mathop{\sum}_{i=1}^{N_s}}m_i(m_i-1).$$ Note that the topological genus is different from the arithmetic genus for algebraic curves with singularities. The ring torus is topologically equivalent to an elliptic curve with no singularities while the spindle torus (see Fig. \ref{fig:topchange}) is equivalent to an elliptic curve with a singularity of multiplicity 2.}} We would like to point out that the topology changing transitions discussed here are similar to the dynamical topology change discussed in \cite{TopChange}. It would be interesting to study more general topology changes where the internal manifold with topological genus-$g$ changes to a geometry with topological genus-$g'$. These topology changing transitions suggest the possible existence of tunneling transitions that cannot be described by conventional Coleman-De Luccia instantons {\cite{Coleman}}. In particular, the lower dimensional effective theory framework used to describe Coleman-De Luccia instantons cannot describe tunneling transitions that involve mixing of an arbitrarily large number of Kaluza-Klein modes.

The family of solutions in (\ref{sol6d}) are free of singularities when $ \text{max} (r(t))< R$, but it is not clear if these solutions are all stable. Since these solutions are obtained from flat solutions in higher dimensions, we expect these solutions to be perturbatively stable. It seems worthwhile to analyze the stability of these solutions.

Another concern that needs to be addressed is the following: How can such solutions be consistent with second law of thermodynamics? The gravitational entropy of the universe reaches a minimum when the universe bounces from a contracting phase to an expanding phase. When the geometry does not admit a closed trapped surface the definition of gravitational entropy is not even clear; in particular, it is not possible to define the gravitational entropy as the area of a Killing horizon. It seems that there exists some notion of time's arrow that can be defined using Raychaudhuri equation even when the universe bounces periodically. The arrow of time defined using the Raychaudhuri equation is related to the the seven-dimensional arrow of time. However, it is not clear if the thermodynamic arrow of time is actually related to this.
 It is also not clear, how quantum effects modify the singularity theorems. So a classical bouncing solution that is geodesically complete and stable could be unstable quantum mechanically.

\section*{Acknowledgments}
We would like to thank M.~Douglas, C.~P.~Herzog, K.~Jensen, S.~Kachru, V.~Kumar, R.~Loganayagam, J.~McGreevy,  E.~\'O~Colg\'ain, D.~S.~Park, N.~Prabhakar, M.~Rocek, and A.~Vikman for discussions, comments and encouragement. We would like to thank M.~Douglas, C.~P.~Herzog, K.~Jensen and J.~McGreevy for comments on the manuscript. 
This work was supported in part by the National Science Foundation under Grants No.\ PHY-0844827 and PHY-1316617.

\begin{appendix}
\section{A short discussion on Kaluza-Klein reduction}
 \setcounter{equation}{0}
\numberwithin{equation}{section}
In this appendix we show that the field configuration in (\ref{sol6d}) is a saddle point of the action in (\ref{EMD}). We show this by showing that 
(\ref{EMD}) can be obtained as a consistent truncation of (\ref{StrinAct}) by setting $H=dB$ and $d\varphi$ to zero. Following is a short description of the proof.

 Let us consider the following ansatz for the seven-dimensional line element, dilaton and the two form field:
$$ ds^{2} = {g}_{MN}dx^{M}dx^{N} = e^{-\sigma/2}{\hat{g}}_{\mu\nu} dx^{\mu}dx^{\nu} + e^{2\sigma} (dz+\hat{A})^{2}$$
\be \tilde{B}_{2} = B_{2} + a_\mu dx^\mu \wedge (dz+\hat{A})  , \quad \varphi = \Phi. \ee  
In the above ansatz, $\tilde{g}$ is the seven-dimensional metric, $\tilde {B}_{2}$ is 
the seven-dimensional two-form field, $\varphi$ is the seven dimensional scalar field and $M, N \in \{t,\vec{x},\theta,\phi,z\}$. 
In order to prove that the six-dimensional field configuration in (\ref{sol6d}) is a solution of the Einstein-Maxwell-Dilaton action in  (\ref{EMD}), we have to prove that the seven dimensional equations of motion truncate to the six-dimensional equations of motion. 
We will now show that the six dimensional equations of motion are satisfied if the seven dimensional equations are satisfied (with $d\Phi =0$ and $H=0$) and vice versa. 
When $d\Phi = 0$ and $H=0$, the seven-dimensional Einstein equations can be written as (see \cite{maharanaschwarz, Scherk} for details of the computation), 
$$ \(\tilde{R}_{\mu\nu} - \half  \tilde{g}_{\mu \nu} \tilde{R}\)  = \(\hat{R}_{\mu\nu} - \half  \hat{g}_{\mu \nu} \hat{R}\)-{5\over 4}\(\nabla_\mu \sigma \nabla_\nu \sigma - \hat{g}_{\mu \nu} \half{\nabla_{\mu} \sigma \nabla^\mu \sigma}\)  $$ \be-{1\over 2} e^{5\sigma/2}\(\hat{F}_{\mu \lambda}\hat{F}^{\lambda}_{\nu} -{1\over 4} \hat{F}^2 \hat{g}_{\mu \nu}\) \label{red1}\ee
\be \(\tilde{R}_{\mu z} - \half  \tilde{g}_{\mu z} \tilde{R}\) = \nabla_\nu \(e^{ 5 \sigma/2} \hat{F}^{\nu\mu}\) = 0 \label{red2}\ee
\be \(\tilde{R}_{zz} - \half  \tilde{g}_{z z} \tilde{R}\) = \nabla_{\mu} \nabla^{\mu} \sigma  = {1\over 4} e^{5\sigma/2} \hat{F}^2 \label{red3}\ee
where all variables decorated by $\sim$ are 7D fields and hatted variables are six-dimensional fields. Equations in (\ref{red1})-(\ref{red3}) are the seven dimensional equations of motion in terms of the six dimensional fields. Note that the equations in (\ref{red1})-(\ref{red3}) are the equations of the motion of the six-dimensional Einstein-Maxwell-Dilaton theory described by (\ref{EMD}).\footnote
{We have made use of the following formula 
\be {1\over \sqrt{-g} }{\delta \over \delta g_{ab}} \int d^{d}x \sqrt{-g}X R = X\(R_{ab} - \half R g_{ab}\) - \nabla_{a}\nabla_{b}X + g_{ab} \nabla_{c}\nabla^{c}X~~ .\ee}
We know that (\ref{CrazyFlat}) is a solution of the seven dimensional equations of motion with $d\Phi =0$ and $H=0$ because it is locally flat. Hence, the field configuration in (\ref{sol6d}) obtained through Kaluza-Klein reduction satisfies the six dimensional equations of motion in (\ref{red1})-(\ref{red3}). We can also show that the six dimensional action in (\ref{EMD}) is a consistent truncation of (\ref{StrinAct}) when $dB_2$, $da$ and $d\Phi$ are zero. To show this we have to show that the equations of motion of $\varphi$ and $\tilde{B}_2$ are also satisfied. Note that the equation of motion for the seven dimensional $B-$ field ($\tilde{B}_2$) is trivial satisfied if $B_{2} $ and $a_\mu$ are zero. The equation fo motion for the seven dimensional scalar field is satisfied (with $d\Phi =0$) iff the following equation is satisfied:
$$ \hat{R} - {5 \over 4} \(\nabla_\mu \sigma\)\(\nabla^\mu \sigma\)  +{3\over 2} \(\nabla_\mu \nabla^\mu \sigma\) - {1\over 4} e^{5 \sigma/2} \hat{F}^2 =0 $$
The above equation is satisfied if equations (\ref{red1}) and (\ref{red2}) are satisfied. In other words, seven dimensional equations of motion are satisfied when the six dimensional equations are satisfied and vice versa. In the next subsection, we will show that the field configuration in (\ref{sol6d}) satisfies the 6D equations of motion by direct substitution.
\subsection{Direct Verification}
First we will show that the gauge field in (\ref{sol6d}) satisfies its equation of motion. To show this we need to compute all non-zero components of $\hat{F}^{\mu \nu}$. 
We can show that only $\hat{F}^{\theta\phi}= -F^{ \phi \theta}$ is non-zero and it is given by
$$ e^{5 \sigma/2}\sqrt{-\hat{g}}\hat{F}^{\theta\phi} = {2 \alpha\beta }$$
The above equation shows that 
$$\d_\mu \(e^{5 \sigma/2}\sqrt{-\hat{g}}\hat{F}^{\theta\phi} \)= 0$$
Hence Maxwell's equation is satisfied  by (\ref{sol6d}).
We will now show that the scalar equation of motion is also satisfied  by (\ref{sol6d}). First we note that,
$$e^{5\sigma/2} \hat{F}^2 = {8 \alpha^2 \beta^2 \over \beta^2 + \alpha^2 g_{\phi \phi}} $$
Evaluating $\nabla_\mu \nabla^\mu \sigma$ we find,
$$ {1\over \sqrt{-\hat{g}}} \d_{\mu} \(\sqrt{-\hat{g}} \hat{g}^{\mu \nu} \d_\nu \sigma \) = {1\over 4} e^{5\sigma/2} \hat{F}^2$$
Hence, the scalar field equation of motion is also satisfied by (\ref{sol6d}). 
We will now show that Einstein's equations are also satisfied. We will first list out all non-zero components of the Einstein tensor, scalar field stress tensor and the gauge field stress tensor. 
\subsubsection{Einstein Tensor}
Einstein tensor is defined as
$\G_{\mu \nu} = \hat{R}_{\mu \nu} - \half \hat{R} \hat{g}_{\mu \nu}$. We will now list all non-zero components of the Einstein tensor. 
\begin{eqnarray}\G_{tt}&=&\frac{\alpha ^2 \left(-10 \alpha ^4 g_{\phi \phi }^2 \left(\cos (2 \theta ) r'(t)^2-1\right)+\alpha ^2 \beta ^2 g_{\phi \phi } \left((21-5 \cos (2 \theta )) r'(t)^2+26\right)\right)}{16 \left(\beta ^2+\alpha ^2 g_{\phi \phi }\right){}^3} \nonumber\\
&~&\qquad\qquad+\frac{\alpha^2 \beta ^4 \left(\sin ^2(\theta ) r'(t)^2+1\right)}{ \left(\beta ^2+\alpha ^2 g_{\phi \phi }\right){}^3}\end{eqnarray}
\begin{eqnarray}
\G_{t\theta}& = &\G_{\theta t} = \frac{\alpha ^2 \sin (\theta ) \cos (\theta ) r(t) r'(t) \left(8 \beta ^4+5 \alpha ^2 g_{\phi \phi } \left(\beta ^2+2 \alpha ^2 g_{\phi \phi }\right)\right)}{8 \left(\beta ^2+\alpha ^2 g_{\phi \phi }\right){}^3}
\end{eqnarray}
\begin{eqnarray}
\G_{t\phi} = \G_{\phi t} = \frac{\alpha ^3 \beta  g_{\phi \phi } \cos (\theta ) r'(t) \left(5 \alpha ^2 g_{\phi \phi }-8 \beta ^2\right)}{8 \left(\beta ^2+\alpha ^2 g_{\phi \phi }\right){}^3}
\end{eqnarray}
\begin{eqnarray}
\G_{\theta\theta} = \frac{\alpha ^2 r(t)^2 \left(16 \beta ^4 \cos ^2(\theta )+\alpha ^2 g_{\phi \phi } \left(21 \beta ^2+5 \cos (2 \theta ) \left(\beta ^2+2 \alpha ^2 g_{\phi \phi }\right)\right)\right)}{16 \left(\beta ^2+\alpha ^2 g_{\phi \phi }\right){}^3}
\end{eqnarray}
\begin{eqnarray}
\G_{\theta\phi} = \G_{\phi\theta} = \frac{\alpha ^3 \beta  g_{\phi \phi } \sin (\theta ) r(t) \left(8 \beta ^2-5 \alpha ^2 g_{\phi \phi }\right)}{8 \left(\beta ^2+\alpha ^2 g_{\phi \phi }\right){}^3}
\end{eqnarray}
\begin{eqnarray}
\G_{\phi\phi} = \frac{\alpha ^2 \beta ^2 g_{\phi \phi } \left(8 \beta ^2-5 \alpha ^2 g_{\phi \phi }\right)}{8 \left(\beta ^2+\alpha ^2 g_{\phi \phi }\right){}^3}
\end{eqnarray}
\begin{eqnarray}
\G_{ij} = -\frac{\alpha ^2 \left(8 \beta ^2+5 \alpha ^2 g_{\phi \phi }\right)}{8 \left(\beta ^2+\alpha ^2 g_{\phi \phi }\right){}^2}\delta_{ij}
\end{eqnarray}

\subsubsection{Scalar Field Stress Tensor} 
The scalar field stress tensor is given by 
$$T^{\text{scalar}}_{\mu\nu} ={5\over 4}\(\nabla_\mu \sigma \nabla_\nu \sigma - \hat{g}_{\mu \nu} \half{\nabla_{\mu} \sigma \nabla^\mu \sigma}\) $$
Following is a list of all non-zero components of the scalar field stress tensor.
\begin{eqnarray}
T^{\text{scalar}}_{tt} = \frac{5 \alpha ^4 g_{\phi \phi } \left(\alpha ^2 g_{\phi \phi } \left(1-\cos (2 \theta ) r'(t)^2\right)+\beta ^2 \left(\sin ^2(\theta ) r'(t)^2+1\right)\right)}{8 \left(\beta ^2+\alpha ^2 g_{\phi \phi }\right){}^3}
\end{eqnarray}
\begin{eqnarray}
T^{\text{scalar}}_{t\theta} =T^{\text{scalar}}_{\theta t} = \frac{5 \alpha ^4 g_{\phi \phi } \sin (\theta ) \cos (\theta ) r(t) r'(t) \left(\beta ^2+2 \alpha ^2 g_{\phi \phi }\right)}{8 \left(\beta ^2+\alpha ^2 g_{\phi \phi }\right){}^3}
\end{eqnarray}
\begin{eqnarray}
T^{\text{scalar}}_{t\phi} = T^{\text{scalar}}_{\phi t} =\frac{5 \alpha ^5 \beta  g_{\phi \phi }^2 \cos (\theta ) r'(t)}{8 \left(\beta ^2+\alpha ^2 g_{\phi \phi }\right){}^3}
\end{eqnarray}
\bea
T^{\text{scalar}}_{\theta\theta} = \frac{5 \alpha ^4 g_{\phi \phi } r(t)^2 \left(\beta ^2 \cos ^2(\theta )+\alpha ^2 g_{\phi \phi } \cos (2 \theta )\right)}{8 \left(\beta ^2+\alpha ^2 g_{\phi \phi }\right){}^3}
\eea
\bea
T^{\text{scalar}}_{\theta\phi} = T^{\text{scalar}}_{\phi\theta} = -\frac{5 \alpha ^5 \beta  g_{\phi \phi }^2 \sin (\theta ) r(t)}{8 \left(\beta ^2+\alpha ^2 g_{\phi \phi }\right){}^3}
\eea
\bea
T^{\text{scalar}}_{\phi\phi} = -\frac{5 \alpha ^4 \beta ^2 g_{\phi \phi }^2}{8 \left(\beta ^2+\alpha ^2 g_{\phi \phi }\right){}^3}
\eea
\bea
T^{\text{scalar}}_{ij} = -\frac{5 \alpha ^4 g_{\phi \phi }}{8 \left(\beta ^2+\alpha ^2 g_{\phi \phi }\right){}^2}\delta_{ij}
\eea

\subsubsection{Gauge Field Tress Tensor}
The gauge field stress tensor is given by 
$$T^{\text{gauge}}_{\mu\nu} ={1\over 2} e^{5\sigma/2}\(\hat{F}_{\mu \lambda}\hat{F}^{\lambda}_{\nu} -{1\over 4} \hat{F}^2 \hat{g}_{\mu \nu}\)  $$
Following is a list of all non-zero components of the gauge field stress tensor.
\bea
T^{\text{gauge}}_{tt} = \frac{\alpha ^2 \beta ^2 \left(\alpha ^2 g_{\phi \phi } \left(r'(t)^2+1\right)+\beta ^2 \left(\sin ^2(\theta ) r'(t)^2+1\right)\right)}{\left(\beta ^2+\alpha ^2 g_{\phi \phi }\right){}^3}
\eea
\bea
T^{\text{gauge}}_{t\theta} = T^{\text{gauge}}_{\theta t} = \frac{\alpha ^2 \beta ^4 \sin (\theta ) \cos (\theta ) r(t) r'(t)}{\left(\beta ^2+\alpha ^2 g_{\phi \phi }\right){}^3}
\eea
\bea
T^{\text{gauge}}_{t\phi} = T^{\text{gauge}}_{\phi t} = -\frac{\alpha ^3 \beta ^3 g_{\phi \phi } \cos (\theta ) r'(t)}{\left(\beta ^2+\alpha ^2 g_{\phi \phi }\right){}^3}
\eea
\bea
T^{\text{gauge}}_{\theta\theta} = \frac{\alpha ^2 \beta ^2 r(t)^2 \left(\beta ^2 \cos ^2(\theta )+\alpha ^2 g_{\phi \phi }\right)}{\left(\beta ^2+\alpha ^2 g_{\phi \phi }\right){}^3}
\eea
\bea
T^{\text{gauge}}_{\theta\phi} = T^{\text{gauge}}_{\phi\theta} = \frac{\alpha ^3 \beta ^3 g_{\phi \phi } \sin (\theta ) r(t)}{\left(\beta ^2+\alpha ^2 g_{\phi \phi }\right){}^3}
\eea
\bea
T^{\text{gauge}}_{\phi\phi}  = \frac{\alpha ^2 \beta ^4 g_{\phi \phi }}{\left(\beta ^2+\alpha ^2 g_{\phi \phi }\right){}^3}
\eea
\bea
T^{\text{gauge}}_{ij} = -\frac{\alpha ^2 \beta ^2}{\left(\beta ^2+\alpha ^2 g_{\phi \phi }\right){}^2}\delta_{ij}
\eea
It is clear from the above expressions that $\G_{\mu \nu} = T^{\text{gauge}}_{\mu\nu}+ T^{\text{scalar}}_{\mu\nu}$. Hence (\ref{sol6d}) satisfies Einstein's equations as well. 
\section{Solutions in Seven Dimensions}
 \setcounter{equation}{0}
\numberwithin{equation}{section}
In this section, we present 7D solutions that are related to the trivial solution in (\ref{CrazyFlat}) by $O(2,2)$ transformations. First, let us look at the solution that can be obtained from (\ref{CrazyFlat}) by using Buscher rules along $z$ direction. We write down the Buscher rules here for convenience: 
\bea
g_{zz}' = {1\over g_{zz}}
~~~~~~&
g_{az}' = \displaystyle{{B_{az}\over g_{zz}}}
~~~~~~&
g_{ab}' = g_{ab} - {g_{az}g_{zb}+B_{az}B_{zb}\over g_{zz}}\\
\varphi' = \varphi -\half\ln{g_{zz}}
~~~~~~&
B_{az}' = \displaystyle{{g_{ay}\over g_{zz}}}
~~~~~~&
B_{ab}' = B_{ab} - {g_{az}B_{zb}+B_{az}g_{zb}\over g_{zz}}
\eea

The 7D solution we get using Buscher transformations is given by
\be ds^2  = e^{-\sigma/2} \hat{g}_{t t} dt^2 + \vec{dx}^2+ 2e^{-\sigma/2} \hat{g}_{t\theta}dt d\theta +e^{-\sigma/2} \hat{g}_{\theta\theta}d\theta^2 + \beta^2 e^{-2\sigma} \(d\phi + A^{(2)}_a dx^a\)^2 +   e^{-2\sigma} dz^2,~ \label{7Dsola}\ee
\be B = \hat{A}_a dx^a \wedge dz + \hat{A}_\phi d\phi \wedge dz, ~\varphi = \varphi_0-\sigma \label{7Dsolb}\ee
where $a\in \{t,\vec{x}\}$ and $A^{(2)}_a = e^{2\sigma} g_{\phi a}$. We can verify that this solution also reduces to (\ref{sol6d}).  
Solutions generated using a general $O(2,2)$ duality transformation on (\ref{sol6d}) need not be equivalent to the above solution. Under general $O(2,2)$ transformations the two dimensional part of the internal manifold and the $B$ field transforms as described in (\ref{OddTrans}). As an example, let us study the action of the following $O(2,2)$ matrix on the 6D solution in (\ref{sol6d})
$$
\Omega ={1\over 2}\left[
\begin{matrix}
1+c & s & c-1 & -s \cr
-s & 1-c & -s & 1+c \cr
c-1 & s & 1+c & -s \cr
s & 1+c & s & 1-c \cr 
\end{matrix}\right]
$$
where $c=\cosh \mu$ and $s =\sinh \mu$ (following the notations in \cite{maharanagaspven}). The internal manifold and $B$ field transforms as follows:
$$\tilde{g}_{\phi \phi} = {(1 +c)^2  +  g_{\phi \phi}^2 + (1-c + s\alpha)^2 \beta^2  + g_{\phi \phi} \(-2(1+c) s\alpha  +(1+c)^2 \alpha^2 + s^2 (1+\beta^4)\)\over 4 g_{\phi \phi} \beta^2 } $$
$$\tilde{g}_{z \phi} = {s \left( -(-1+c-2s \alpha + \alpha^2 + s)g_{\phi \phi}   - (1+c) \beta^2\)  (1 + g_{\phi \phi} \beta^2)\over 4 g_{\phi \phi} \beta^2 } $$
$$\tilde{g}_{zz} = {s^2\beta^2  +  g_{\phi \phi}^2 + (s -(1+c)\alpha)^2 \beta^2  + g_{\phi \phi} \(-2c(1+ s \alpha-\beta^4) +1 + 2s\alpha +s^2 \alpha^2 + c^2 (1+\beta^4)\)\over 4 g_{\phi \phi} \beta^2 }  $$
$$\tilde{B}_{z \phi} = {\(\left( g_{\phi \phi} (-1 + c - 2 s \alpha + (1+c) \alpha^2) + (1+c)\beta^2\)\(-g_{\phi\phi}(s - 2 c \alpha + s \alpha^2) -s\beta^2 \)\)\over 4 g_{\phi \phi} \beta^2} $$
We can verify that this solution reduces to a solution that is not equivalent to (\ref{sol6d}).
Note that the above solution and the solution in (\ref{7Dsola}, \ref{7Dsolb}) can be uplifted to solutions of type II supergravity trivially. 

\end{appendix}


\begin{thebibliography}{100}

 \bibitem{GuthBordeVilenkin} 
  A.~Borde, A.~H.~Guth and A.~Vilenkin,
  ``Inflationary space-times are incompletein past directions,''
  Phys.\ Rev.\ Lett.\  {\bf 90}, 151301 (2003)
  [gr-qc/0110012].
  
   \bibitem{Hawking}
  S.~W.~Hawking and G.~F.~R.~Ellis, {\it The Large scale
   structure of space-time}, Cambridge University Press, Cambridge(1973)

  \bibitem{Bouncing}
  R.~R.~Caldwell,
  ``A Phantom menace?,''
  Phys.\ Lett.\ B {\bf 545}, 23 (2002)
  [astro-ph/9908168]. 
  R.~H.~Brandenberger,
  ``The Matter Bounce Alternative to Inflationary Cosmology,''
  arXiv:1206.4196 [astro-ph.CO]. (review)
\bibitem{ghost}
N. ~Arkani-Hamed, H. ~C. ~Cheng, M. ~A. ~Luty and S. ~Mukohyama, 
``Ghost condensation and a consistent infrared modification of gravity," 
JHEP 0405, 074 (2004), [hep-th/0312099].
P. ~Creminelli, M. ~A. ~Luty, Al. ~Nicolis, L. ~Senatore, 
``Starting the Universe: Stable Violation of the Null Energy Condition and Non-standard Cosmologies," 
JHEP 0612, 080 (2006), [hep-th/0606090].
\bibitem{Modified} 
  S.~Alexander, T.~Biswas and R.~H.~Brandenberger,
  ``On the Transfer of Adiabatic Fluctuations through a Nonsingular Cosmological Bounce,''
  arXiv:0707.4679 [hep-th].
  T.~Biswas, A.~Mazumdar and W.~Siegel,
  ``Bouncing universes in string-inspired gravity,''
  JCAP {\bf 0603}, 009 (2006)
  [hep-th/0508194].
    \bibitem{SHU} 
  P.~W.~Graham, B.~Horn, S.~Kachru, S.~Rajendran and G.~Torroba,
  ``A Simple Harmonic Universe,''
  arXiv:1109.0282 [hep-th].
  \bibitem{Vikman1} 
  I.~Sawicki and A.~Vikman,
  ``Hidden Negative Energies in Strongly Accelerated Universes,''
  Phys.\ Rev.\ D {\bf 87}, no. 6, 067301 (2013)
  [arXiv:1209.2961 [astro-ph.CO]].
\bibitem{Dubovsky} 
  S.~Dubovsky, T.~Gregoire, A.~Nicolis and R.~Rattazzi,
  ``Null energy condition and superluminal propagation,''
  JHEP {\bf 0603}, 025 (2006)
  [hep-th/0512260].
\bibitem{Vikman2} 
  D.~A.~Easson, I.~Sawicki and A.~Vikman,
  ``When Matter Matters,''
  JCAP {\bf 1307}, 014 (2013)
  [arXiv:1304.3903 [hep-th], arXiv:1304.3903].
  \bibitem{Vikman3} 
  E.~Babichev, V.~Mukhanov and A.~Vikman,
  ``k-Essence, superluminal propagation, causality and emergent geometry,''
  JHEP {\bf 0802}, 101 (2008)
  [arXiv:0708.0561 [hep-th]].

\bibitem{BrandVafa} 
  R.~H.~Brandenberger and C.~Vafa,
  ``Superstrings in the Early Universe,''
  Nucl.\ Phys.\ B {\bf 316}, 391 (1989).
  
\bibitem{Toumbas} 
  C.~Kounnas and N.~Toumbas,
  ``Aspects of String Cosmology,''
  PoS Corfu {\bf 2012}, 083 (2013)
  [arXiv:1305.2809 [hep-th]].
  
  \bibitem{vengasperini} 
M.~Gasperini and G.~Veneziano, Astropart. Phys. {\bf 1} (1993) 317.
  M.~Gasperini and G.~Veneziano,
  ``The Pre - big bang scenario in string cosmology,''
  Phys.\ Rept.\  {\bf 373}, 1 (2003)
  [hep-th/0207130].
G.~Veneziano, ``String cosmology: The Pre - big bang scenario", hep-th/0002094.
   G.~Veneziano, Phys. Lett. B {\bf 265} (1991) 287;  
  K.~A.~Meissner and G.~Veneziano,
  ``Symmetries of cosmological superstring vacua,''
  Phys.\ Lett.\ B {\bf 267}, 33 (1991).
 
  \bibitem{maharanagaspven} 
  M.~Gasperini, J.~Maharana and G.~Veneziano,
  ``From trivial to nontrivial conformal string backgrounds via $O(d,d)$ transformations,''
  Phys.\ Lett.\ B {\bf 272}, 277 (1991).
   K.~A.~Meissner and G.~Veneziano,
  ``Manifestly $O(d,d)$ invariant approach to space-time dependent string vacua,''
  Mod.\ Phys.\ Lett.\ A {\bf 6}, 3397 (1991)
  [hep-th/9110004].
   M.~Gasperini and G.~Veneziano,
  ``$O(d,d)$ covariant string cosmology,''
  Phys.\ Lett.\ B {\bf 277}, 256 (1992)
  [hep-th/9112044].
  M.~Gasperini, J.~Maharana and G.~Veneziano,
  ``Boosting away singularities from conformal string backgrounds,''
  Phys.\ Lett.\ B {\bf 296}, 51 (1992)
  [hep-th/9209052].
  \bibitem{KOSST} 
  J.~Khoury, B.~A.~Ovrut, N.~Seiberg, P.~J.~Steinhardt and N.~Turok,
  ``From big crunch to big bang,''
  Phys.\ Rev.\ D {\bf 65}, 086007 (2002)
  [hep-th/0108187].
  \bibitem{Seiberg} 
  N.~Seiberg,
  ``From big crunch to big bang: Is it possible?,''
  [hep-th/0201039].
  \bibitem{GibbHorTown}
  G.~W.~Gibbons, G.~T.~Horowitz and P.~K.~Townsend,
  ``Higher dimensional resolution of dilatonic black hole singularities,''
  Class.\ Quant.\ Grav.\  {\bf 12}, 297 (1995)
  [hep-th/9410073].
    \bibitem{Behrndt} 
  K.~Behrndt and S.~Forste,
  ``String Kaluza-Klein cosmology,''
  Nucl.\ Phys.\ B {\bf 430}, 441 (1994)
  [hep-th/9403179].
  \bibitem{larsenWilczek} 
  F.~Larsen and F.~Wilczek,
  ``Resolution of cosmological singularities,''
  Phys.\ Rev.\ D {\bf 55}, 4591 (1997)
  \bibitem{Feinstein} 
  A.~Feinstein and M.~A.~Vazquez-Mozo,
  ``M theory resolution of four-dimensional cosmological singularities,''
  Nucl.\ Phys.\ B {\bf 568}, 405 (2000)
  [hep-th/9906006].
\bibitem{KOST} J.~Khoury, B.A.~Ovrut,
P.J.~Steinhardt and N.~Turok, [hep-th/0103239].
 \bibitem{Orbifold} 
  G.~T.~Horowitz and A.~R.~Steif,
  ``Singular string solutions with nonsingular initial data,''
  Phys.\ Lett.\ B {\bf 258}, 91 (1991).
  H.~Liu, G.~W.~Moore and N.~Seiberg,
  ``Strings in a time dependent orbifold,''
  JHEP {\bf 0206}, 045 (2002)
  [hep-th/0204168].
   \bibitem{john} 
  M.~Fabinger and J.~McGreevy,
  ``On smooth time dependent orbifolds and null singularities,''
  JHEP {\bf 0306}, 042 (2003)
  [hep-th/0206196].
  
\bibitem{LMS2} 
  H.~Liu, G.~W.~Moore and N.~Seiberg,
  ``Strings in time dependent orbifolds,''
  JHEP {\bf 0210}, 031 (2002)
  [hep-th/0206182].
  \bibitem{HoroPol} 
  G.~T.~Horowitz and J.~Polchinski,
  ``Instability of space - like and null orbifold singularities,''
  Phys.\ Rev.\ D {\bf 66}, 103512 (2002)
  [hep-th/0206228].
\bibitem{SenovillaDust} 
  J.~M.~M.~Senovilla,
  ``New class of inhomogeneous cosmological perfect-fluid solutions without big-bang singularity,''
  Phys.\ Rev.\ Lett.\  {\bf 64}, 2219 (1990).

    \bibitem{Traschen} 
  D.~Kastor and J.~Traschen,
  ``Magnetic Fields in an Expanding Universe,''
  arXiv:1312.4923 [hep-th].
\bibitem{Buscher} 
  T.~H.~Buscher,
  ``A Symmetry of the String Background Field Equations,''
  Phys.\ Lett.\ B {\bf 194}, 59 (1987).

\bibitem{Garriga} 
  J.~Garriga,
  ``Smooth ``creation" of an open universe in five dimensions,''
  [hep-th/9804106].
  
  \bibitem{Melvin} 
  M.~A.~Melvin,
  ``Pure magnetic and electric geons,''
  Phys.\ Lett.\  {\bf 8}, 65 (1964).
  
  \bibitem{Peet} 
  A.~W.~Peet,
  ``TASI lectures on black holes in string theory,''
  hep-th/0008241.

\bibitem{SenBH}
  A.~Sen,
  ``Twisted black p-brane solutions in string theory,''
  Phys.\ Lett.\ B {\bf 274}, 34 (1992)
  [hep-th/9108011].

\bibitem{HorBH} 
  G.~T.~Horowitz,
  ``The dark side of string theory: Black holes and black strings.,''
  [hep-th/9210119].
   
\bibitem{maharanaschwarz} 
  J.~Maharana and J.~H.~Schwarz,
  ``Noncompact symmetries in string theory,''
  Nucl.\ Phys.\ B {\bf 390}, 3 (1993)
  [hep-th/9207016].
\bibitem{BK73} 
  V.~A.~Belinski and I.~M.~Khalatnikov,
  ``Effect of Scalar and Vector Fields on the Nature of the Cosmological Singularity,''
  Sov.\ Phys.\ JETP {\bf 36}, 591 (1973).

\bibitem{OlafBarton} 
  O.~Hohm, D.~Lust and B.~Zwiebach,
  ``The Spacetime of Double Field Theory: Review, Remarks, and Outlook,''
  arXiv:1309.2977 [hep-th].
  
 \bibitem{Geroch} 
  R.~P.~Geroch,
  ``What is a singularity in general relativity?,''
  Annals Phys.\  {\bf 48}, 526 (1968).
  
  \bibitem{CKR} 
  B.~Craps, D.~Kutasov and G.~Rajesh,
  ``String propagation in the presence of cosmological singularities,''
  JHEP {\bf 0206}, 053 (2002)
  [hep-th/0205101].
  
  \bibitem{ABKS} 
  O.~Aharony, M.~Berkooz, D.~Kutasov and N.~Seiberg,
  ``Linear dilatons, NS five-branes and holography,''
  JHEP {\bf 9810}, 004 (1998)
  [hep-th/9808149].
  
  \bibitem{Qlattices} 
  A.~Donos and J.~P.~Gauntlett,
  ``Holographic Q-lattices,''
  arXiv:1311.3292 [hep-th].
  
  
  
\bibitem{TrapSurf}
  ~J.~M.~M.~Senovilla, 
  ``Trapped surfaces, horizons and exact solutions in higher dimensions,''
  Class.\ Quant.\ Grav. 19, L113 (2002)
  arXiv:0204005v2  [hep-th].
  ~J.~M.~M.~Senovilla, 
  ``Trapped surfaces,''
  arXiv:1107.1344v2  [gr-qc].


\bibitem{TopChange}
E.~Kiritsis and C.~Kounnas,
  ``Dynamical topology change, compactification and waves in a stringy early universe,''
  [hep-th/9407005].

\bibitem{TWF}
  ~S. ~Giddings, ~A. ~Maharana, 
  ``Dynamics of warped compactifications and the shape of the warped landscape,''
   Phys.\ Rev.\ D {\bf 73}, 126003 (2006)
   arXiv:hep-th/0507158v4.
~M. ~R. ~Douglas, ~G. ~Torroba, 
  ``Dynamics of warped compactifications and the shape of the warped landscape,''
   JHEP {\bf 0905}, 013 (2009)
   arXiv:hep-th/0805.3700.
  G.~Shiu, G.~Torroba, B.~Underwood and M.~R.~Douglas,
  ``Dynamics of Warped Flux Compactifications,''
  JHEP {\bf 0806}, 024 (2008)
  [arXiv:0803.3068 [hep-th]].
\bibitem{Coleman} 
  S.~R.~Coleman and F.~De Luccia,
  ``Gravitational Effects on and of Vacuum Decay,''
  Phys.\ Rev.\ D {\bf 21}, 3305 (1980).
  %
\bibitem{Scherk} 
  J.~Scherk and J.~H.~Schwarz,
  ``How to Get Masses from Extra Dimensions,''
  Nucl.\ Phys.\ B {\bf 153}, 61 (1979)
\end{thebibliography}

\end{document}